\documentclass[%
 reprint,
 amsmath,amssymb,
 aps,
prb,
]{revtex4-2}

\usepackage[utf8]{inputenc}
\usepackage{graphicx}
\usepackage{dcolumn}
\usepackage{bm}
\usepackage{amsfonts}
\usepackage{color}

\begin{document}

\preprint{APS/123-QED}

\title{Transport in the 2D Fermi-Hubbard Model: Lessons from Weak Coupling}

\author{Thomas G. Kiely}
 \altaffiliation{tgk37@cornell.edu}
\author{Erich J. Mueller}
 \altaffiliation{em256@cornell.edu}
\affiliation{Laboratory of Atomic and Solid State Physics, Cornell University, Ithaca, NY 14853}

\date{\today}

\begin{abstract}
We use quantum kinetic theory to calculate the  thermoelectric transport properties of the 2D single band Fermi-Hubbard model in the weak coupling limit. For generic filling, we find that the high-temperature limiting behaviors 
of the electrical ($\sim T$) and thermal ($\sim T^2$) resistivities 
persist down to temperatures of order the hopping matrix element
$T\sim t$, almost an order of magnitude below the bandwidth. 
At half filling, perfect nesting leads to anomalous low temperature scattering and nearly  $T$-linear electrical resistivity at all temperatures.  
We hypothesize that the $T$-linear resistivity observed in recent cold atom experiments is 
continuously connected to this weak coupling physics and suggest avenues for experimental verification.
We find a number of other novel thermoelectric results, such as a low-temperature Wiedemann-Franz law with Lorenz coefficient $5\pi^2/36$.  
\end{abstract}

\maketitle

\section{\label{sec:intro}Introduction}
One of the most significant open problems in condensed matter physics is the origin of 
``strange metal" behavior in strongly correlated materials. 
This 
non-Fermi-liquid
behavior is often identified experimentally through anomalous transport properties: a DC resistivity which is $T$-linear down to low temperatures and a mean-free path which becomes shorter than the lattice spacing at high temperatures.  The latter is referred to as a violation of the the Mott-Ioffe-Regel (MIR) limit. These observations differ from 
expected
Fermi liquid behavior, which is characterized by a resistivity that is proportional to $T^2$ at low temperatures and a saturation of the MIR bound at high temperatures~\cite{coleman,mott,ioffe}. Such anomalous behaviors have been observed in a diverse array of strongly-correlated materials~\cite{Hussey,ruthenate,pnictide,heavyFermion,univPlanck,scattRt} and have invited a variety of sophisticated theoretical and numerical approaches to explain them~\cite{cuprateNS,critpt,holographic,highTPerspective,hartnollBadMetal,numericalHubbardVertex,Huang,slopeInvariantTLinear}.

One longstanding difficulty of studying these materials directly is the complex interplay of interactions between electrons, phonons, and impurities. 
For this reason, there has been considerable interest in the cold atom community to simulate non-trivial strongly-correlated model systems. The Fermi-Hubbard model is a natural starting point for these investigations, as cold atoms in an optical lattice naturally realize a nearest-neighbor hopping model with onsite interactions. The former is controlled using the lattice depth, while the latter is tuned via a Feshbach resonance. Early experimental evidence from the Bakr group indicates that the 2D realization of this model has a high-temperature strange metal phase~\cite{Brown379}.  This conclusion is supported by advanced numerical methods~\cite{numericalHubbardVertex,Huang,slopeInvariantTLinear} and analytic high-temperature expansions~\cite{highTPerspective}. Of particular note, the authors find that the Nernst-Einstein decomposition of the conductivity, $\sigma=D\chi$, does not shed light onto the origin of this behavior: both the diffusion constant and the charge compressibility have non-trivial temperature dependences in the strange metal regime, conspiring to give a $T$-linear resistivity. Furthermore, the diffusion constant appears to saturate a high-temperature bound that would be conceptually consistent with the MIR limit~\cite{Brown379}.

In this paper we 
clarify this story by studying weak-coupling transport in the 2D Fermi Hubbard model.
We use a quantum kinetic theory to show that, even at weak coupling, the resistivity is nearly $T$-linear down to temperatures which are an order of magnitude below the bandwidth. At temperatures which are large compared to the bandwidth, this behavior is attributed to a vanishing inverse effective mass,  arising  from competing contributions from both the top and bottom of the band.  Despite the diverging resistivity, the quasiparticle scattering rate in this regime saturates at an interaction-dependent value that is well below the MIR bound.  
%
Remarkably, 
at intermediate temperatures,
$T$-linearity arises
from a non-trivial interplay between the effective mass and the scattering lifetime. This is analogous to the aforementioned ``conspiracy" between the compressibility and diffusion constant seen in experiments.
We demonstrate that $T$-linearity persists to arbitrarily low temperatures in the vicinity of half filling, where the density of states diverges and the Fermi surface is perfectly nested.
Away from half-filling we find the
the conventional $T^2$ behavior at sufficiently low temperature, with a crossover to $T$-linearity at higher temperature.  The crossover temperature vanishes at half-filling and for small Fermi surfaces, $|\mu_F|>2t$, where umklapp scattering is forbidden. 

Low-temperature $T$-linear resistivity 
is seen in a number of theoretical studies of
the 2D Fermi-Hubbard model~\cite{nestedFL,fujimoto,schlottmann2,leeMFL} and 
other models with van-Hove singularities, such as twisted bilayer graphene~\cite{tbg}. These references draw interesting connections to marginal Fermi liquid theory~\cite{MarginalFLTheory}. Our paper uses elementary arguments to show that these features can be found at weak coupling and explains the behavior in terms of kinetic theory and perfect nesting. 

Cold atom systems differ from conventional materials in that they have defect-free lattices that do not support phonons.
Given that transport properties in most materials are dominated by phonons or impurities, this puts us in a novel transport regime:
current dissipation in cold atom realizations of the Fermi-Hubbard model arises only from the analog of electron-electron scattering.  At low temperature the dominant processes involve umklapp scattering, where the lattice absorbs momentum corresponding to a reciprocal lattice vector \cite{peierls}.  
 
Interactions in cold atom experiments are tuned by changing the lattice potential or by using a Feshbach resonance \cite{expReview1,expReview2}.  
The recent Bakr experiments were conducted in a moderate coupling limit where our kinetic theory is not expected to be quantitatively accurate.  
Nonetheless, 
our results capture the qualitative behavior.
Further experiments at weaker coupling would 
quantitatively test our results.

Beyond calculating the conductivity $\sigma$, we explore thermal conductivity $\kappa$ and more general thermoelectric properties.  At low temperatures we find that the Lorenz number $L=\kappa/(\sigma T)$ approaches a constant.  This Wiedemann-Franz law~\cite{wiedemannFranz} is expected when the same degrees of freedom are responsible for thermal and electrical transport.  Our Lorenz number, however, differs from what is found in a system where the dissipation is dominated by impurity scattering.  
The Wiedemann-Franz law breaks down at high temperature.

We organize our paper as follows. In Section~\ref{sec:model} we discuss the 2D Fermi-Hubbard model as well as the variational approach which we use to solve the Boltzmann equation \cite{Ziman}. In Section~\ref{sec:results} we present our results, divided between the electrical properties (Sec.~\ref{subsec:Resistivity}) and the full thermoelectric matrix (Sec.~\ref{subsec:ThermoEl}). We discuss the experimental implications of our work in Section~\ref{sec:exptApp}, and in Section~\ref{sec:conclusion} we summarize our conclusions.

\section{\label{sec:model}Model}
In this paper we study the 2D single-band Fermi-Hubbard model on a rectangular lattice with nearest-neighbor hopping:
\begin{equation}
    \mathcal{H}=-t\sum_{\langle ij\rangle,\sigma}(c^\dagger_{i\sigma}c_{j\sigma}+h.c.)+U\sum_ic^\dagger_{i\uparrow}c^\dagger_{i\downarrow}c_{i\downarrow}c_{i\uparrow},
\end{equation}
where $c^{(\dagger)}_{i\sigma}$ is the fermionic annihilation (creation) operator on site $i$ with spin $\sigma$. In condensed matter systems, this model describes highly localized orbitals with an onsite interaction parameterized by $U$. It is a natural model Hamiltonian for fermionic cold atoms in an optical lattice, where the interaction strength, $U$, is tuned using a Feshbach resonance and the tunneling strength, $t$, is set by the optical lattice depth.

Although we are largely thinking of cold atom realizations, we use the language of electronic systems.  We interpret a force ${\bf F}$ in terms of an electric field ${\bf E}={\bf F}/e$.  The charge current is simply the number current times the electron charge, ${\bf j}=e{\bf j_n}$.

\subsection{\label{subsec:lbe}Linearized Boltzmann Equation}
The richness of the Hubbard model arises from the non-commutativity of the kinetic and interaction terms: the kinetic term is diagonal in momentum space, with a dispersion $\epsilon_k=-2t\cos(k_x)-2t\cos(k_y)$, while the interaction term is diagonal in real space. Our paper will study this model in the weak-coupling regime, $U \ll t$, such that the interaction 
may be treated as a perturbation. In this limit the excitations are quasiparticle states with well-defined crystal momenta. 
The interaction term introduces collisions between quasiparticles, giving them a finite lifetime. 
Fermi liquid theory holds that there is a domain of finite $U/t$ within which this analysis is valid. 

We model the fermion distribution function, $f_k(r)$, which counts how many quasiparticles of a given spin state are at position $r$ with momentum $\hbar k$.  The  particle density is $n(r)={ 2}\int\frac{d^2k}{(2\pi)^2}f_k(r)$, where the factor of 2 accounts for spin.  The distribution function obeys 
 a Boltzmann equation,
\begin{equation}
    \partial_tf_k+\nabla_rf_k\cdot {\bf v}_k + e{\bf E}\cdot\nabla_kf_k=I_k[f],
    \label{eq:boltzeq}
\end{equation}
where ${\bf v}_k=\frac{1}{\hbar}\nabla_k \epsilon_k$ is the velocity.
The collision integral, $I_k[f]$, is a functional that determines the rate at which particles scatter into and out of the momentum state $k$. 
It can be calculated with 
Fermi's Golden Rule:
\begin{equation}
    I_k[f]=-\frac{2\pi}{\hbar}\sum_{if}p_i \big|\langle f|\mathcal{H}_{int}|i\rangle\big|^2\big(n_k^{(f)}-n_k^{(i)}\big)\delta(\epsilon_f-\epsilon_i)
    \label{fgr}
\end{equation}
where $|i\rangle$ and $|f\rangle$ are Slater determinants, $p_i$ is the probability of initially being in state $|i\rangle$,  $\mathcal{H}_{int}=U\sum_i c_{i\uparrow}^\dagger c_i^\uparrow c_{i\downarrow}^\dagger c_{i\downarrow}$,  
$n_k^{(i)}=\langle i|c^\dagger_{k\sigma}c_{k\sigma}|i\rangle$, and
$f_k=\langle c^\dagger_{k\sigma}c_{k\sigma}\rangle=\sum_i p_i n_k^{(i)}$. Replacing $\mathcal{H}_{int}$ with the full many-body T-matrix gives a formally exact value for the decay rate; 
Eq.~(\ref{fgr})
corresponds to the Born approximation, where one keeps only the leading-order term after expanding in powers of $U/t$. 

If one assumes that the momentum states are are uncorrelated, the collision integral is given by
\begin{eqnarray}
    I_k[f]&=&\frac{2\pi U^2}{\hbar}\!\!\!\!\sum_{k',k'',k''',Q}
    \!\!\!\!\!\delta_{k+k'-k''-k'''-Q}
    ~\delta(\epsilon_f-\epsilon_i)\times\\\nonumber
    && \big(f^{\prime\prime} f^{\prime\prime\prime} (1-f)(1-f^\prime) - f f^\prime (1-f^{\prime\prime})(1-f^{\prime\prime\prime})\big)
\end{eqnarray}
where we have used the short-hand notation $f=f_k,f^\prime=f_{k^\prime}$, and so on.
We have explicitly included the sum over reciprocal lattice vectors, $Q$, which accounts for momentum non-conserving umklapp scattering events. Note that the integrand is exactly zero for $f_k=f^0_k(r)$, the Fermi-Dirac distribution:
\begin{equation}
f^0_k(r)=\frac{1}{e^{\beta(r)(\epsilon_k-\mu(r))}+1}.
\label{eq:fkr}
\end{equation}
We take $\beta(r)=1/k_B T(r)$ and $\mu(r)$ to be slowly varying, treating $\nabla \beta$, $\nabla \mu$, and $\bf E$ as small parameters.

We linearize the Boltzmann equation by taking $f_k-f^0_k=-\Phi_k\frac{\partial f^0_k}{\partial \epsilon_k}$, where $\Phi_k$ is formally small. 
We can always choose $\beta$ and $\mu$ so that this perturbation does not change the density or energy,
$\int\frac{d^2k}{(2\pi)^2}\Phi_k\frac{\partial f^0_k}{\partial \epsilon_k}=\int\frac{d^2k}{(2\pi)^2}(\epsilon_k-\mu)\Phi_k\frac{\partial f^0_k}{\partial \epsilon_k}=0$. The linearized collision integral, in the thermodynamic limit, is given by
\begin{widetext}
\begin{multline}
I_k[\Phi]= -\frac{\Lambda\beta}{(2\pi)^3}\sum_Q\int d^2k'\int d^2k''\int d^2k'''(\Phi_k+\Phi_{k'}-\Phi_{k''}-\Phi_{k'''})f^0_{k}f^0_{k'}(1-f^0_{k''})(1-f^0_{k'''}) \times \\ \times \delta^{2}(k+k'-k''-k'''-Q)\delta(\epsilon_k+\epsilon_{k'}-\epsilon_{k''}-\epsilon_{k'''})
\label{eq:colInt}
\end{multline}
\end{widetext}
where $\Lambda=\frac{U^2a^4}{\hbar}$ and $a$ is the lattice spacing.

\subsection{\label{subsec:varSol}Variational Solution}
The thermoelectric matrix is obtained from the steady-state solutions to the Boltzmann equation, where $\partial_tf_k=0$. The resulting equation is an inhomogeneous integral equation for $\Phi_k$. We follow the procedure set out in Ref.~\cite{Ziman} to obtain a variational bound on the transport coefficients. 
We make the ansatz
$\Phi_k=\sum_i\xi_i\phi^{(i)}_k$,
where $\phi^{(i)}_k$ are a fixed set of trial functions.
The goal will be to determine the optimal set of coefficients, $\{\xi_i\}$, such that the resulting distribution is as close to the actual Boltzmann equation solution as possible.  In our numerical calculations we will use a two-term ansatz, with $\phi^{(1)}_k=(\nabla_k\epsilon_k)_x$ and $\phi^{(2)}_k=(\epsilon_k-\mu)(\nabla_k\epsilon_k)_x$, though in this section we consider the completely general case.  The theory becomes exact in the limit where the $\phi_k^{(i)}$ form a complete set.

We  define the particle and heat currents arising from each trial function, respectively, as
\begin{equation}
    \begin{aligned}
        j^{(i)}_\alpha&=-{ 2e}\int\frac{d^2k}{(2\pi)^2}(\nabla_k\epsilon_k)_\alpha\phi^{(i)}_k\frac{\partial f^0_k}{\partial \epsilon_k} \\u^{(i)}_\alpha&=-{ 2}\int\frac{d^2k}{(2\pi)^2}(\epsilon_k-\mu)(\nabla_k\epsilon_k)_\alpha\phi^{(i)}_k\frac{\partial f^0_k}{\partial \epsilon_k}
    \end{aligned}
    \label{eq:currents}
\end{equation}
where $\alpha=x,y,z$. These currents are generated by the electric field, ${\bf E}$, included explicitly in Eq.~(\ref{eq:boltzeq}), as well as a spatially-homogeneous temperature gradient, $\nabla_r T$. The latter force comes from the spatial derivative of $\beta(r)$ in the second term of Eq.~(\ref{eq:boltzeq}). 
Gradients of $\mu(r)$  play the same role as the electric field, and
we follow the standard condensed matter convention of defining an effective field $E+(1/e)\nabla_r\mu$ that generates particle currents~\cite{Ashcroft}. In what follows, we will use the variable $E$ to denote this combination of an external field and the gradient of the chemical potential. Furthermore, we will neglect the effect of density gradients on the steady-state properties of the system. In this particular problem, neglecting density gradients can be justified by noting that 
the Hartree terms which couple density gradients to currents are of
subleading order in $U/t$. 
Furthermore, we envision a current-carrying state of constant density.

Linearizing the Boltzmann equation, multiplying by $\Phi_k$, and integrating over $k$ yields
%
\begin{equation}
    \sum_i\xi_i\bigg(\frac{j^{(i)}_\alpha E_\alpha}{T}+u^{(i)}_\alpha\nabla_\alpha\bigg(\frac{1}{T}\bigg)\bigg)=\frac{1}{T}\sum_{ij}\xi_i\xi_jP_{ij}
    \label{eq:entEOM}
\end{equation}
where
\begin{widetext}
\begin{multline}
    P_{ij} =  \frac{\Lambda\beta}{(2\pi)^5}\sum_Q\int d^2k\int d^2k'\int d^2k''\int d^2k'''\big(\phi^{(i)}_k+\phi^{(i)}_{k'}-\phi^{(i)}_{k''}-\phi^{(i)}_{k'''}\big)\big(\phi^{(j)}_k+\phi^{(j)}_{k'}-\phi^{(j)}_{k''}-\phi^{(j)}_{k'''}\big)\times \\\times f^0_{k}f^0_{k'}(1-f^0_{k''})(1-f^0_{k'''}) \delta(k+k'-k''-k'''-Q)\delta(\epsilon_k+\epsilon_{k'}-\epsilon_{k''}-\epsilon_{k'''}).
    \label{eq:pij}
\end{multline}
\end{widetext}
Under the assumption that the forces are small, the $f_k^0$ can be taken as homogeneous in this expression.
Eq.~(\ref{eq:entEOM}) does not uniquely define the set $\{\xi_i\}$. Onsager \cite{onsager1,onsager2} argued that the optimal choice of $\{\xi_i\}$ is the one that maximizes the rate of entropy production from scattering.
Appendix~\ref{sec:eom}, modeled after Ref.~\cite{Ziman2}, gives an explicit derivation in the present context. 
The three terms in Eq.~(\ref{eq:entEOM}) represent the rates of entropy change from the external field, temperature gradient, and scattering.
$\dot{S}_{scatter}=-\dot{S}_{field}-\dot{S}_{inhom}$.
Following the optimization procedure in Appendix~\ref{sec:var},
we find
\begin{equation}
\xi_i = \sum_j(P^{-1})_{ij} 
\bigg(\frac{j^{(j)}_\alpha E_\alpha}{T}+u^{(j)}_\alpha\nabla_\alpha\bigg(\frac{1}{T}\bigg)\bigg).
\label{eq:optxi}
\end{equation}
We define the thermoelectric matrix, following Ref.~\cite{Ziman}, as
\begin{equation}
    \begin{pmatrix} J \\ U \end{pmatrix}=L\begin{pmatrix}
    E \\ \nabla T
    \end{pmatrix},
\end{equation}
where $J_\alpha=\sum_i\xi_ij^{(i)}_\alpha$ and $U_\alpha=\sum_i\xi_iu^{(i)}_\alpha$ are the total number and heat currents. Inserting Eq.~(\ref{eq:optxi}) into these definitions yields
\begin{equation}
    L=\begin{pmatrix} \sum_{ij}j^{(i)}_\alpha(P^{-1})_{ij}j^{(j)}_\beta & -\frac{1}{T}\sum_{ij}j^{(i)}_\alpha(P^{-1})_{ij}u^{(j)}_\beta \\ \sum_{ij}u^{(i)}_\alpha(P^{-1})_{ij}j^{(j)}_\beta & -\frac{1}{T}\sum_{ij}u^{(i)}_\alpha(P^{-1})_{ij}u^{(j)}_\beta \end{pmatrix}.
    \label{eq:thermoelectricMat}
\end{equation}
One can determine a variety of transport coefficients in terms of the components of the thermoelectric matrix. In this paper, we report the DC charge and thermal resistivities ($\rho=1/\sigma$ and $1/\kappa$, respectively), the Seebeck coefficient ($\alpha$), and the Lorenz number ($L_0=\frac{\kappa}{T\sigma}$):
\begin{equation}
    \begin{split}
        &\rho=\frac{1}{L_{11}} \hspace{1cm} 1/\kappa=-\frac{L_{11}}{\det(L)} \\&\alpha=-\frac{L_{12}}{L_{11}} \hspace{1cm} L_0=-\frac{1}{T}\frac{\det(L)}{L_{11}^2}
        \label{eq:thermDef}
    \end{split}
\end{equation}

One can readily verify that the entropy-maximizing condition produces an upper bound on 
$\rho$ and the \textit{bare} thermal resistivity, $1/\bar{\kappa}=1/L_{22}$.  The latter corresponds to the thermal response for $E=0$, as opposed to the more physical condition $J=0$.
The other coefficients, $\alpha$, $\kappa$, and $L_0$,
do not necessarily satisfy a variational bound. 

\section{\label{sec:results}Results}

We use two trial functions in the variational calculation: $\phi^{(1)}_k=(\nabla_k\epsilon_k)_x$ and $\phi^{(2)}_k=(\epsilon_k-\mu)(\nabla_k\epsilon_k)_x$. These are natural deviations from equilibrium to generate charge ($\phi^{(1)}$) and heat ($\phi^{(2)}$) currents.  In Appendix~\ref{sec:valid} we estimate that the resulting low-temperature resistivities are accurate to within $30\%$. 
We expect similar accuracy at high temperature.

We divide our results between electrical and thermal properties in Sec.~\ref{subsec:Resistivity} and \ref{subsec:ThermoEl}, respectively.

\subsection{\label{subsec:Resistivity}Resistivity and Scattering Rate}

\begin{figure}[tb]
    \centering
    \includegraphics[width=3.375in]{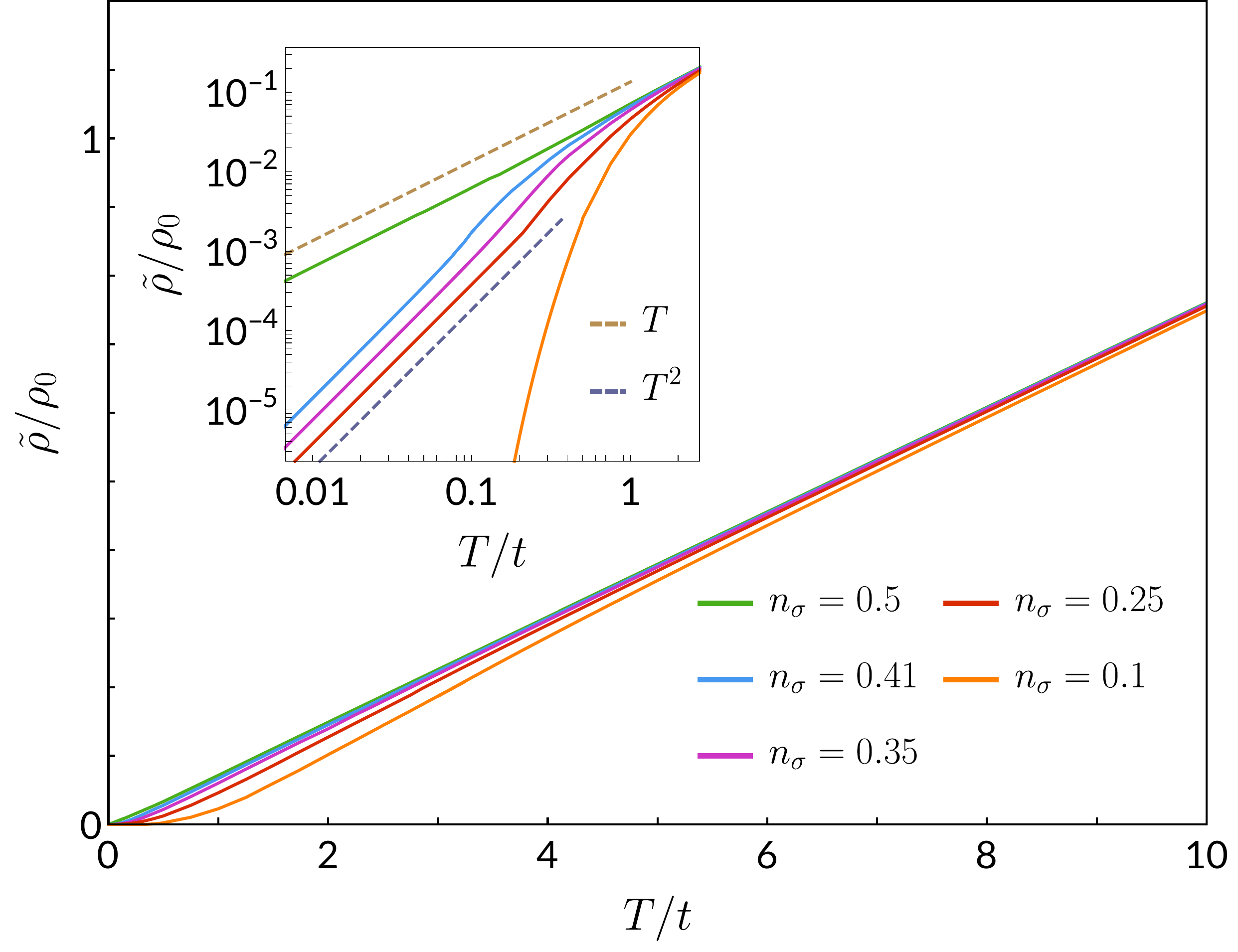}
    \caption{(color online) Rescaled resistivity, $\tilde{\rho}=(t/U)^2~\rho$, versus temperature, scaled by $\rho_0=e^2/\hbar$. Data is shown for a variety of densities: $n_\sigma=$ 0.5 (green), 0.41 (blue), 0.35 (purple), 0.25 (red), and 0.1 (orange). The resistivity is a monotonic increasing function of temperature for all densities with a $T$-linear high temperature asymptote, $\tilde{\rho}_\infty/{\rho}_0\approx0.076~T/t$. This asymptote is approached most quickly for densities near half filling.  Inset: Zoom in to low temperatures on a log-log scale, showing the crossover to $T^2$ behavior. At $n_\sigma\leq0.185$ the Fermi surface is sufficiently small that no umklapp processes are possible at zero temperature, so the resistivity decays exponentially. At $n_\sigma=0.5$, perfect nesting leads to asymptotic $T$-linear resistivity down to zero temperature. Power law guides to the eye are given by the dashed lines.
    }
    \label{fig:resFig}
\end{figure}

Figure~\ref{fig:resFig} shows the  resistivity of the 2D Fermi-Hubbard model due to quasiparticle-quasiparticle scattering, calculated by numerically performing the integrals from Sec.~\ref{subsec:varSol}.  In our weak-coupling picture, the only $U$-dependence comes from the fact that the scattering rate (and hence the resistivity) is proportional to $(U/t)^2$.  
%
We find that the resistivity is a monotonic increasing function of temperature that vanishes at $T=0$. These are the hallmarks of metallic behavior. As will be  explained in Sec.~\ref{subsubsec:HighTRes}, the high-temperature asymptotic behavior is linear in temperature and independent of the particle density. Next-leading-order high-temperature corrections are of order $1/T$ and are minimized at half filling. The high temperature $T$-linear behavior persists to surprisingly low temperature, and it would require very high-precision experiments to identify the deviations for $T\gtrsim t$.  The deviations from linear are particularly small at half-filling, though they are non-zero.

As seen in the inset of Fig.~\ref{fig:resFig}, at low temperature there are three different behaviors, depending on the filling.  For $0<|\mu_F|<2t$ we find $\rho\propto T^2$, while for $\mu_F=0$ we instead find $\rho\propto T$.  In Sec.~\ref{subsubsec:LowTRes} we explain this difference in terms of band structure.  For $|\mu_F|>2t$, low temperature umklapp scattering is forbidden and the resistivity falls off exponentially. 

\begin{figure}[ht]
    \centering
    \includegraphics[width=3.375in]{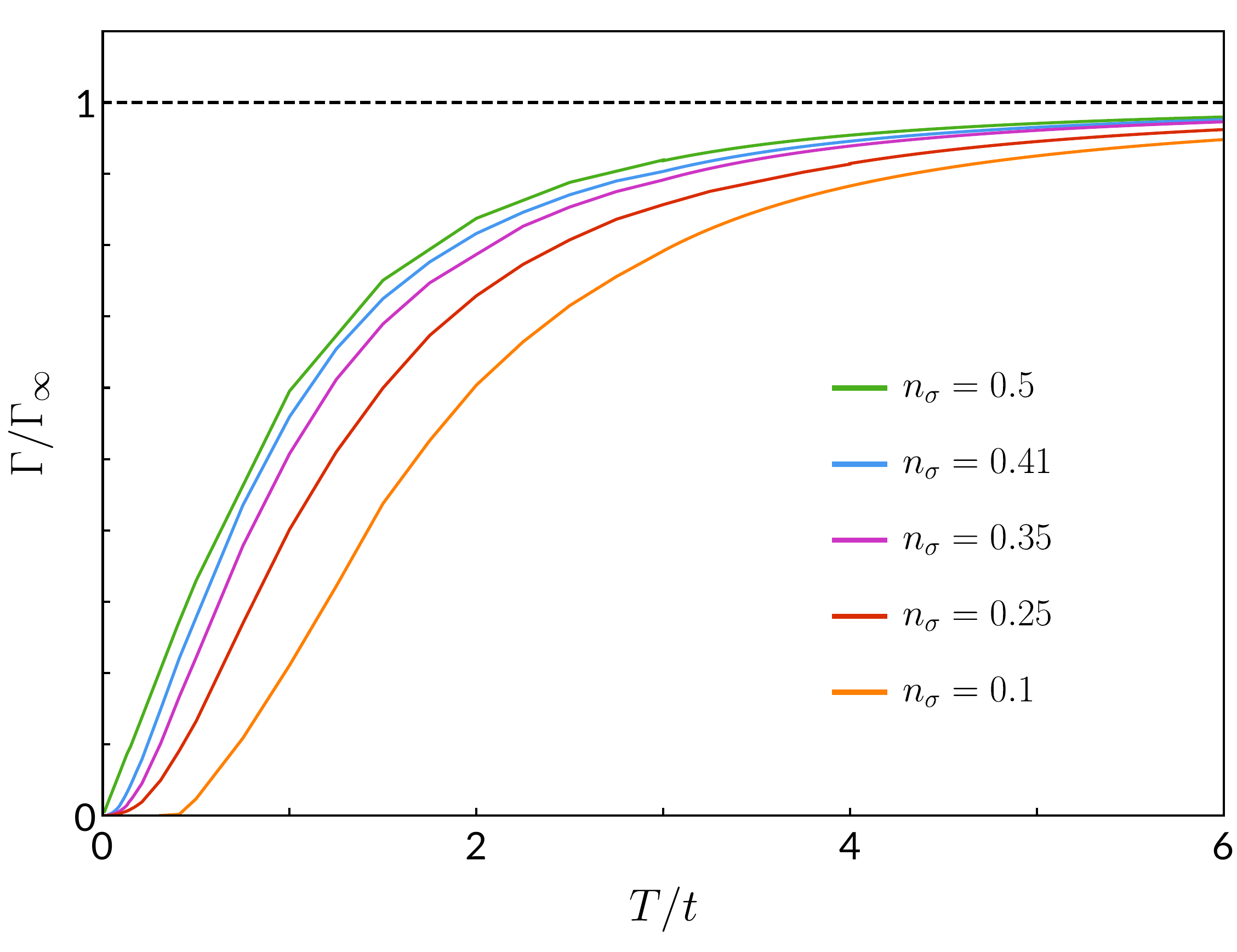}
    \caption{(color online) Scattering rate, $\Gamma$, versus temperature in units of the high-temperature asymptote, $\Gamma_\infty\approx0.609~n_\sigma(1-n_\sigma)~U^2/\hbar t$. 
    At low temperatures, $\Gamma\propto T^2$ for $0<|\mu_F|<2t$; the scattering rate vanishes as $\sim T$ at half filling and exponentially for $|\mu_F|>2t$.}
    \label{fig:gamFig}
\end{figure}

We define a scattering lifetime using the Einstein relation, $\sigma=D\chi_c$, and the definition of the diffusion constant in a quasiparticle system, $D=\frac{1}{d}\langle v^2\rangle \tau$ ($d$ is the number of spatial dimensions). It is straightforward to compute $\langle v^2\rangle$, the average squared quasiparticle velocity, and $\chi_c$, the charge compressibility, for the non-interacting gas. The resulting scattering rate, $\Gamma=1/\tau$, is plotted in Fig.~\ref{fig:gamFig}.
In the limit of infinite temperature, the scattering rate saturates.  At low temperature we again find three regimes: $\Gamma\propto T^2$, $T$, and $e^{-\Delta_U/T}$ for $0<|\mu_F|<2 t$, $\mu_F=0$, and $|\mu_F|>2 t$, respectively. Here $\Delta_U=2(|\mu_F|-2t)$ is the umklapp gap~\cite{umklappGap}.

\subsubsection{\label{subsubsec:HighTRes}High Temperature}
Cold atom experiments measuring transport in the 2D Fermi-Hubbard model have thus far been limited to moderate-to-high temperatures, $T/t\gtrsim 1$~\cite{Brown379,demarco}. In this section we model this regime. 

At high temperature we can expand the Fermi functions as $f^0_k=n_\sigma-n_\sigma(1-n_\sigma)\beta\epsilon_k$.  It is then straightforward to write a high temperature series expansion for the integrals in Eq.~(\ref{eq:currents}) and (\ref{eq:pij}).  We find  $\tilde{\rho}_\infty(T)=0.076~(T/t) \rho_0$, where $\rho_0=e^2/\hbar$.  Similarly ${\Gamma}_\infty=0.609~n_\sigma(1-n_\sigma)~U^2/\hbar t$.
A useful way to interpret these asymptotic results  is in terms of a diverging effective mass within a Drude picture, where $\sigma=\frac{ne^2 \tau}{m^*}$ with $\tau=1/\Gamma$. There are positive and negative contributions to the inverse effective mass from the bottom and top of the band.  These cancel at high temperatures, resulting in a divergent resistivity despite the fact that the scattering rate saturates.
More precisely, in a relaxation time approximation
\begin{equation}
\bigg(\frac{n}{m^*}\bigg)_{\rm eff}=\int \frac{d^2k}{2\pi^2} f_k (\nabla_k^2 \epsilon_k),
\end{equation}
and for large $T$ this integral vanishes 
as $1/T$.


%
We further interpret the scattering rate as
 $\Gamma=a^{-2}n_\sigma(1-n_\sigma) \sigma_{\rm eff} \bar v$, where $n_\sigma$ is the dimensionless filling fraction and $a$ is the lattice constant.
Since the occupations $f_k$ approach a constant as $T\to\infty$, 
the average velocity approaches $\bar v=\sqrt{\langle v^2\rangle}\to2at/\hbar$. 
This implies that the effective cross-section is $\sigma_{\rm eff}=0.3 a (U/t)^2$.  Up to the numerical prefactor, this last result can be derived from dimensional analysis and the  Born approximation expression $\sigma_{\rm eff}\propto U^2$.

While this calculation is only justified for perturbatively small $U/t$, it 
leads to arbitrarily large resistivity at sufficiently high temperature.  This high temperature divergence is well documented for single-band models \cite{highTPerspective}, and
we emphasize that this should not be interpreted as a violation of the MIR limit: the scattering rate and mean free path remain bounded. 

\subsubsection{\label{subsubsec:LowTRes}Low Temperature}

Low-temperature quantities depend only on properties of the Fermi surface, and are derived by performing a Sommerfeld expansion~\cite{coleman}.
Applying this expansion to Eq.~(\ref{eq:pij}) gives a leading-order $T^2$ behavior of the resistivity for generic filling (Appendix~\ref{sec:lowTform}) and $T$-linear behavior at half-filling (Appendix~\ref{sec:tLinArg}).
The key features of this argument are described below, as well as a more qualitative argument.
%
%
Only momentum non-conserving umklapp processes contribute to the resistivity at low temperature.  When $|\mu_F|>2t$, these processes are geometrically disallowed and the resistivity is exponentially small in $\Delta_U/k_BT$.

To reach a qualitative understanding of this behavior, we consider the rate at which a particle of momentum $k_1$ undergoes scattering.  In particular, we consider processes 
$(k_1,k_2)\rightarrow(k_3,k_4)$, with energies $\epsilon_1,\epsilon_2,\epsilon_3,\epsilon_4$, with $|\epsilon_j-\mu_F|< k_B T$. At low temperature, the number of allowed choices of $k_2$ scale with $T$.  Having fixed $k_1$ and $k_2$, energy and momentum conservation constrains three of the four degrees of freedom of $k_3$ and $k_4$.  One therefore expects that the number of allowed final states should scale as $T$.  Consequently, the scattering rate (and resistivity) scale as $T^2$.  One factor of $T$ is associated with the freedom to choose $k_2$ and the other factor corresponds to redistributing energy between $k_3$ and $k_4$.

At half filling the counting is slightly different. Up to logarithmic corrections from the divergent density of states, the number of allowed values of $k_2$ again scales as $T$.  Energy and momentum conservation again restrict all but one degree of freedom of $k_3$ and $k_4$. At half filling, however, the phase space for scattering is dominated by nested scattering events that are automatically within $k_BT$ of the Fermi surface. Thus the number of final states is independent of temperature and the scattering rate scales as $T$.




In Appendix~\ref{sec:tLinArg} 
we put this argument on stronger mathematical foundations.
We express the resistivity as an integral over the energy of pairs of scattering particles, and expand the Fermi functions to arrive at 
\begin{equation}
  \rho\propto  \beta\int_{-8t}^{8t}dE~\frac{(E/2-\mu_F)^2}{\sinh^2(\beta(E/2-\mu_F))}f_T(E).
    \label{eq:naiveP11}
\end{equation}
Up to numerical factors, $T^2 f_T(E)$ is the joint density of states for scattering, restricting the particle energies to be within $k_B T$ of $E/2$.  It is well approximated by
\begin{equation}
    f_T(E)=\begin{cases} \frac{1}{16\pi^4}\sqrt{\big(\frac{4t}{E}\big)^2-1} & |E|>cT \\ \frac{1}{16\pi^4}\sqrt{\big(\frac{4t}{cT}\big)^2-1} & |E|\leq cT \end{cases},
    \label{eq:fT}
\end{equation}
where $c$ is a numerical constant. As long as $|\mu_F|\neq 0$, we can take the limit
\begin{equation}
\lim_{T\to 0} f_T(E)\equiv f(E) = \frac{1}{16\pi^4} \sqrt{(4t/E)^2-1}.
\end{equation}
The first term in the integrand of Eq.~(\ref{eq:naiveP11}) becomes a delta-function as $T\to 0$, and we recover the expected $T^2$ resistivity.  At $\mu_F=0$, however, $f_T(E)\propto 1/T$ and the resistivity is $T$-linear.
For $|\mu_F|>2t$ the resistivity vanishes
as there are no allowed umklapp processes.

\begin{figure}
    \centering
    \includegraphics[width=3.375in]{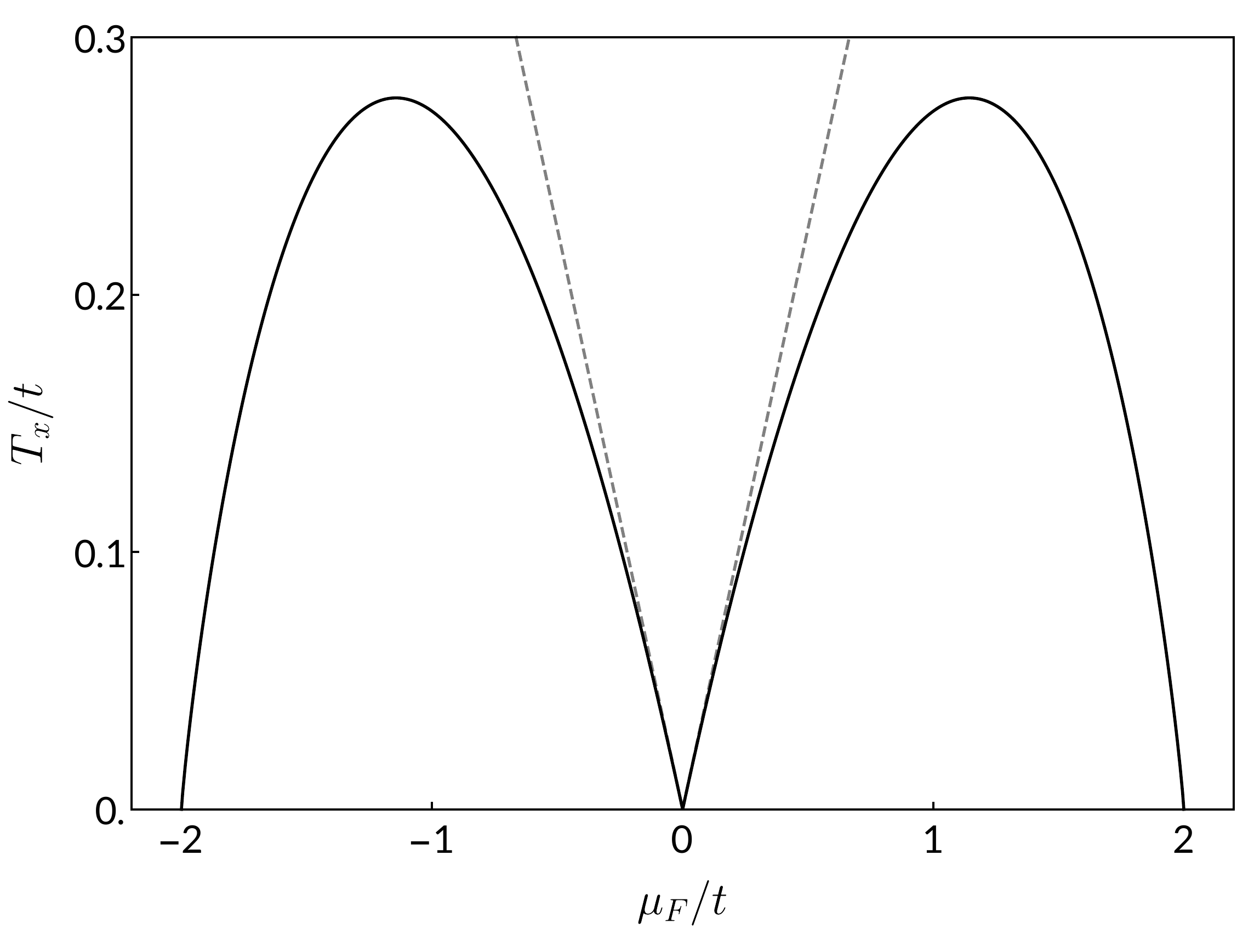}
    \caption{Crossover temperature, $T_x$, where the dominant $T^2$ term in the Sommerfeld expansion of the resistivity is equal to the exponentially-suppressed subleading correction (black). Near half filling we see that the crossover temperature vanishes as $T_x\propto |\mu_F|$ (grey dashed line). Umklapp scattering is geometrically forbidden for small Fermi surfaces, which causes $T_x$ to vanish as $|\mu_F|\to 2t$.
    }
    \label{fig:xtemps}
\end{figure}

At generic filling, the strong $E=0$ peak in $f_T(E)$ gives a subleading contribution to the resistivity which scales as $ T e^{-2 |\mu_F|/T}$.  Thus one has a crossover between a low-temperature $T^2$ regime and a higher-temperature linear-$T$ behavior.  Figure~\ref{fig:xtemps} illustrates this crossover by finding the temperature, $T_x$, where this sub-leading term is equal to the dominant $T^2$ contribution.  This crossover is also evident in the full numerical results in Fig.~\ref{fig:resFig}.

The crossover temperature vanishes as $|\mu_F|\to2t$ due to the geometric exclusion of umklapp processes.
Near half filling, $T_x$ vanishes as $\sim|\mu|$ and it is natural to interpret the crossover in terms of the thermal occupation of the nested $E=0$ states. At half filling, the crossover temperature vanishes.  It is noteworthy that $T_x$ is never larger than $0.3 t$, which is an order of magnitude below the bandwidth.

\subsection{\label{subsec:ThermoEl}Thermoelectric Properties}
\begin{figure}
    \centering
    \includegraphics[width=3.375in]{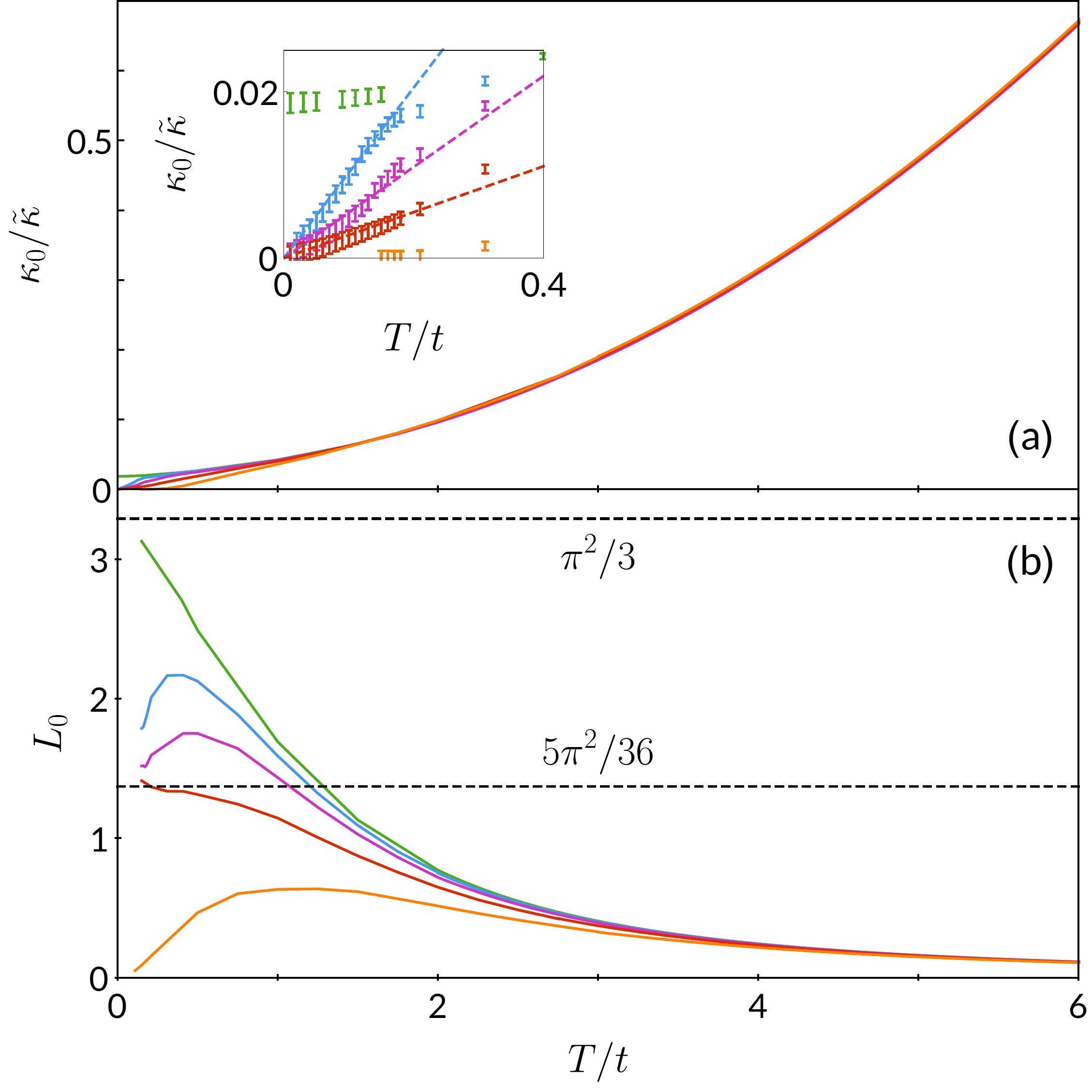}
    \caption{(color online) { (a)} Scaled thermal resistivity, $1/\tilde \kappa=(t/U)^2(1/\kappa)$, in units of $1/\kappa_0=1/\hbar t$, as a function of temperature for a variety of fillings.  See Fig.~\ref{fig:resFig} for key. The thermal resistivity diverges as $T^2$ at high temperatures with a small vertical offset at finite doping. Inset: The thermal resistivity at low temperatures (dots) with error bars from estimated numerical uncertainty. Dotted lines give low-temperature expansion for $0<|\mu_F|<2t$; $1/\kappa$ vanishes exponentially for small Fermi surfaces ($|\mu_F|>2t$). At half filling, the thermal resitivity approaches a constant value of $1/\tilde\kappa\to0.019/{\kappa}_0$. { (b)} Lorenz number, $L_0=\kappa/T\sigma$, versus temperature. The Lorenz number vanishes as $1/T^2$ at high temperatures. A Wiedemann-Franz law is satisfied at zero temperature for $0<|\mu_F|<2t$ with a Lorenz number of $5\pi^2/36$ and at half filling with a Lorenz number that appears to approach $\pi^2/3$ (labeled dashed lines); for $|\mu_F|>2t$, the Lorenz number vanishes at low temperatures.}
    \label{fig:kappaLorenz}
\end{figure}

Figure~\ref{fig:kappaLorenz}
shows the 
thermal resistivity, $1/\kappa$, and the Lorenz number, $L_0=\kappa/T\sigma$, calculated using the techniques described in Sec.~\ref{sec:model}.
The thermal resistivity, plotted in units of $1/\kappa_0=1/\hbar t$, diverges as $T^2$ at high temperatures with a coefficient that is independent of the density: $\tilde{\kappa}_0/\tilde{\kappa}_\infty\approx 0.018~(T/t)^2$. Next-leading-order corrections give a small density-dependent vertical offset that vanishes at half filling.  This high-temperature behavior can be modeled by the same techniques as in Sec.~\ref{subsubsec:HighTRes}.

At low temperatures we find that $1/\kappa$ vanishes linearly in temperature for $0<|\mu_F|<2t$.
At half filling the thermal resistivity approaches a constant ($1/\kappa\to 0.019/{\kappa}_0$)
%
due to the same nesting argument as found in Sec.~\ref{subsubsec:LowTRes}.
%
For small Fermi surfaces ($|\mu_F|>2t$),  umklapp processes are gapped out and  $1/\kappa$ vanishes exponentially. 

Comparing the temperature dependence of the thermal resistivity $\kappa$ and the conductivity $\sigma$, we see that the Lorenz number $L_0=\kappa/T\sigma$ approaches a constant at low temperature.  This behavior is familiar from conventional materials, where elastic impurity scattering leads to $L_0^{\rm elastic}=\pi^2/3$ at low temperatures.  This Wiedemann-Franz relation is an indication that the same mechanism governs thermal and charge diffusion.  It is 
often used as a means to judge the relative elasticity of resistive scattering events~\cite{dasSarmaWF}. 

\begin{figure}
    \centering
    \includegraphics[width=3.375in]{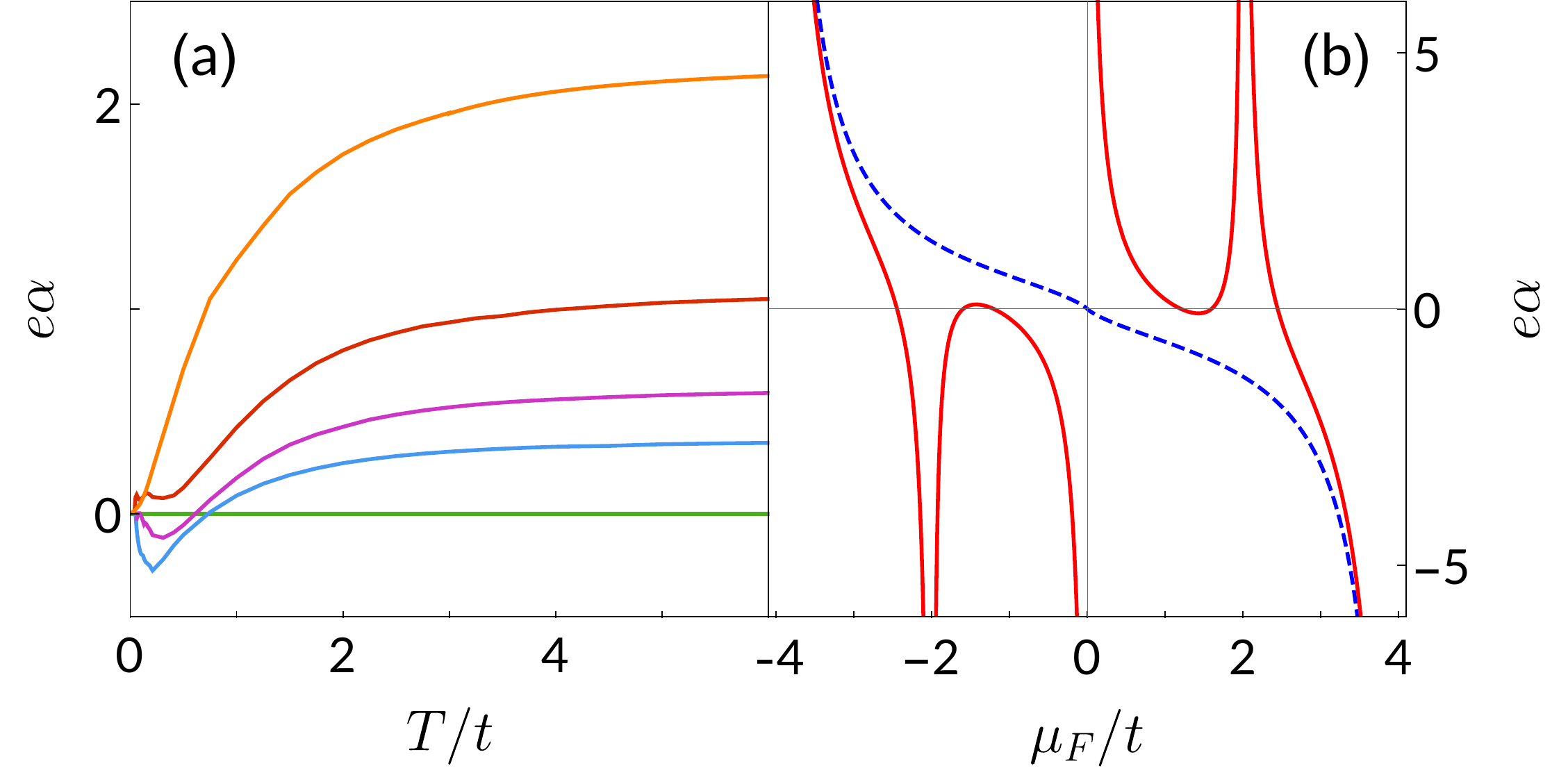}
    \caption{(color online) { (a)} Seebeck coefficient, $\alpha$, versus temperature $T$. At infinite temperatures, the Seebeck coefficient approaches the high temperature free Fermi gas value $e\alpha_\infty=-\ln((1-n_\sigma)/n_\sigma)$. At low temperatures the Seebeck coefficient vanishes linearly with temperature.The Seebeck coefficient is particle-hole antisymmetric and is therefore strictly zero at half filling. { (b)} Slope of the leading-order low-temperature behavior of the Seebeck coefficient versus chemical potential (red) as well as the slope of the free Fermi gas Seebeck coefficient at zero temperature (blue). Introducing scattering causes the slope to diverge at half filling and at $|\mu_F|=2t$. Exponentially-suppressed umklapp scattering cause the deviation between the curves for $|\mu|>2t$.}
    \label{fig:seebeckCoef}
\end{figure}

Through an expansion of the $P_{ij}$ scattering integrals at low temperature, we find that for $0<|\mu_F|<2t$ the low temperature Lorenz number is $L_0=5\pi^2/36$.  This is somewhat smaller than the value coming from elastic impurity scattering.
Figure~\ref{fig:kappaLorenz} confirms 
this result.  At
half filling the Lorenz number appears to approach a different value,
$\pi^2/3$. This is indicative of a qualitative change in the scattering processes. 
For $|\mu_F|>2t$ the Lorenz number vanishes at low temperature. At high temperatures, the Lorenz number vanishes as $1/T^2$: $\kappa\propto T^{-2}$, $\sigma\propto T^{-1}$.
The Seebeck coefficient, or thermopower, $\alpha$, characterizes the voltage induced by a thermal gradient.  It is more complicated to understand than either the electrical and thermal resistivities.  For example, it can have quite rich density dependence \cite{thermopower}.
Figure~\ref{fig:seebeckCoef}(a) shows the temperature dependence of $\alpha$.
At high temperature it approaches the
infinite-temperature non-interacting value, $\alpha_\infty=-\log((1-n_\sigma)/n_\sigma)$, which is derived in Appendix~\ref{sec:freeFermiSeebeck}.
This form is consistent with the 
Heikes formula
\cite{Heikes,chaikin}. At low temperatures the thermopower vanishes linearly in $T$ for all fillings. 

In Eq.~(\ref{eq:thermDef}) we argue that $\alpha=-L_{12}/L_{11}$ where $L_{ij}$ involves a moment of the collision integral.  Within the Born approximation, both $L_{12}$ and $L_{11}$ scale as $U^{-2}$, and hence $\alpha$ is independent of the interaction strength.  Despite its independence from $U$, the $\alpha$ in Eq.~(\ref{eq:thermDef}) differs from that of the non-interacting Fermi gas,
indicating that the $U\to 0$ limit is singular.
Behavior in this regime is often understood in terms of the Mott formula~\cite{Ashcroft},
\begin{equation}
    \alpha_{Mott}=T\frac{\pi^2}{3}\frac{d}{d\mu_F}\ln\big(\rho(\mu_F)\langle\tau(\epsilon,k)\nabla_k\epsilon_k\rangle_{\mu_F}\big).
\end{equation}
The density of states is given by $\rho(x)$ and $\langle\ldots\rangle_{\mu_F}$ denotes momentum averaging over the Fermi surface. The low-temperature slope of the Seebeck coefficient in Fig.~\ref{fig:seebeckCoef}(b) diverges at half filling and at $|\mu_F|=2t$ due to divergences in the log-derivative of the scattering lifetime at those points. It should be noted, however, that the radius of convergence of the low-temperature expansion vanishes at both of those points: beyond $T\approx t$, the effects of these low-temperature divergences are minimal.
Umklapp scattering is exponentially suppressed when $|\mu_F|>2t$, but nonetheless $\alpha$ vanishes linearly in $T$ with a coefficient that differs from that of the ideal gas.

\section{\label{sec:exptApp}Experimental Implications}
To date there have been three cold atom experiments that measure the conductivity of the Fermi-Hubbard model.  The first two experiments, by the Thywissen \cite{thywissen} and DeMarco \cite{demarco} groups, explored  3D Fermi-Hubbard transport.
The Thywissen group
applied a time-varying force to a harmonically trapped lattice gas and measuring the center-of-mass response \cite{thywissen}. They extracted $\sigma(\omega)=\langle j(\omega)\rangle/F(\omega)$, yielding a low frequency conductivity and a transport scattering rate. 
The DeMarco group instead used a Raman pulse to generate spin currents in a 3D Fermi-Hubbard system \cite{demarco}.  From the subsequent decay of these currents they were able to extract a transport lifetime and define a resistivity.  These two experiments are complementary in that one worked in the frequency domain, and the other in the temporal domain.  

The third cold atom experiment, by the Bakr group \cite{Brown379}, involved a 2D lattice.  They
used an additional superlattice potential to create a 
charge-density wave.  After turning off the superlattice, they  imaged the decay of the
density wave.  By repeating the experiment with different wave-vectors, they extracted a diffusion constant and scattering rate. They also measured the charge compressibility, $\chi_c$, and used
the Nernst-Einstein equation  to
infer the conductivity,
$\sigma_{DC}=D\chi_c$.
This experiment serves as the primary point of comparison for our calculations.

The top-line result of the Bakr experiment is that they see a $T$-linear resistivity, which bears a resemblance to the phenomenology of 
``strange-metallic" behavior in correlated metals~\cite{Hussey,ruthenate,pnictide,heavyFermion,univPlanck,scattRt}. This behavior persisted down to temperatures $T/t\lesssim 1$ despite non-trivial temperature dependence in the diffusion constant and compressibility. They also determine that the scattering rate saturates at high temperatures and exhibits a sharp downturn below $T/t\sim 4$. While our weak-coupling calculation do not quantitatively reproduce their results (as they have $U/t\approx 8$), we have demonstrated that all qualitative features are present in the weak-coupling model. On the basis of these observations, we hypothesize that our results are continuously connected to their experiments. 

The clearest test of this hypothesis would be to repeat the Bakr study with weaker interactions.  
This regime could be achieved by tuning the lattice depth, transverse confinement, or  atomic scattering length (via a Feshbach resonance).
One technical challenge with the weakly interacting limit is that, to avoid boundary effects, the atomic cloud must be large compared to the mean free path.  For current experiments, with sizes of order 30 lattice spacings, this restricts $U\gtrsim0.6 t$ at $T/t=0.5$.

An important aspect of our study is the  crossover between the low temperature $T^2$ and high temperature $T^1$ resistivity.  This crossover occurs at temperatures well below those studied in Ref.~\cite{Brown379}. 
In addition to the challenges of achieving these temperatures, reliable low temperature thermometry requires novel approaches~\cite{zwierleinThermometry}.  The crossover temperature is greatest near fillings of $n_\sigma =0.185$ and $0.815$, 
where the umklapp gap opens up.

Nesting plays an important role in our  weakly-interacting transport calculation.  This physics can be explored by adding lattice anisotropy or a superlattice, both of which shift the filling at which nesting occurs. One can also study other lattices which do not display nesting~\cite{computationalPerspective,triang,felipe}.



The density dependence of the resistivity is at least as interesting as the temperature dependence. 
In particular, the most dramatic manifestation of strong-coupling physics is that at half-filling the Fermi Hubbard model describes an interaction-driven insulator: when $T\lesssim U/10$, the resistivity rises as the temperature is reduced~\cite{2DMIT_1,2DMIT_2}.  
When $U$ is large compared to $t$, 
one expects that proximity to this Mott physics will lead to density dependence of the resistivity which significantly differs from our weak-coupling results,
even at intermediate temperatures
~\cite{highTPerspective}.



In addition to calculating the electrical resistivity, we construct the full
 thermoelectric matrix, which also describes heat transport and thermoelectric effects. Measuring thermal transport in cold atoms is quite challenging, but there has been at least one successful experiment \cite{coldAtomHeatEngine,esslinger2}. 
 There 
 the authors used a gate beam to separate two cold atom ``reservoirs" with a quasi-2D channel.   One reservoir is excited, and the temperature of the both reservoirs is monitored.  The thermal conductivity of the channel can them be deduced.  One could imagine adding a lattice to this setup to measure the thermal conductivity of the Fermi Hubbard model.     There may further be approaches based on tilted lattices which give access to the full thermoelectric matrix \cite{guardadoSanchez}.


A more conventional approach to thermoelectric measurements of the Hubbard model might be achieved in transition metal dichacogenide (TMD) heterobilayers, which realize 2D Fermi-Hubbard physics on a Moire lattice~\cite{fai}. Such a scheme would be advantageous insofar as conventional methods for thermoelectric transport could be used. One might tune the effective interaction strength by changing the distance between the gates and the sample: when the gates are closer, they more effectively screen the long-range Coulomb interaction. In general, however, the downside of TMDs compared to cold atoms is in the relative difficulty of tuning the interaction strength as well as the presence of a long-range interaction that complicates the theoretical analysis.  Additionally, lattice defects and phonons may contribute to the resistivity.

\section{\label{sec:conclusion}Conclusions and Outlook}

Experimental studies of cold atom transport in optical lattices are still in their relative infancy. The primary experimental papers cited here have all been published in the last three years, and their full impact has yet to be felt.
Our paper approaches the transport problem from the weak-coupling side, in which calculations are tractable and the physical principles are readily extracted. 

The key conclusion of our study is that even weak coupling models 
can host a variety of ``unconventional" transport properties.
As has been well established in prior work
\cite{highTPerspective}, the high-temperature resistivity diverges in a single-band model.  
This divergence is not associated with a short mean-free path, but rather with a diverging effective mass.  
For all coupling strengths
the the transport coefficients are simple power laws
$\rho\sim T$, $1/\kappa\sim T^2$, $\alpha\sim T^0$.
The prefactors 
have
non-trivial dependence $U$ and $n_\sigma$.  Mapping out this dependence on interactions and density is a prime target for future experiments.



At weak coupling, these high-temperature
results persist to
temperatures on the order of $T\approx t$ (or lower near half-filling). While these are high temperatures in the context of condensed matter systems, it bears re-emphasizing that cold atom experiments have yet to probe transport at temperatures considerably colder than this. 

At moderate temperatures, $1<T/t<4$, we find that the regime of near-$T$-linearity in the electrical resistivity is accompanied by a non-trivial order-of-magnitude decrease of the scattering rate.
In our calculation, the featurelessness of the resistivity in this range of temperatures arises from an interplay between the quasiparticle scattering rate and the effective mass (or, equivalently, between the diffusion constant and the charge compressibility) that is entirely explicable in terms of band theory. We emphasize this point to draw a comparison to a similar phenomenon observed in the Bakr experiment~\cite{Brown379}, which probed the strongly-interacting limit.

At low temperature, we use a Sommerfeld expansion to recover the expected Fermi-liquid result, $\rho\propto T^2$, and similar expressions for the full thermoelectric matrix.  The radius of convergence of this expansion is finite, and it 
%
vanishes 
at half filling, where $\rho\propto T$. 
This anomalous scaling arises from the continuum of umklapp scattering events enabled by the nested bandstructure.  The nesting condition can also lead to various spin-density wave and charge-density wave instabilities which  may preempt some of this behavior \cite{nestedFL,computationalPerspective}.  

The prime motivator of the atomic Hubbard model experiments is trying to gain understanding of strongly correlated phenomena, including high temperature superconductivity.  Such insight will require much lower temperatures.  The pseudogap regime in the Cuprates occurs for $T\lesssim 0.1 t$.  Strange metal behavior is also apparent at those scales.  The crossover between the weak-coupling physics explored in this paper and the strong-coupling physics seen in materials is likely to be quite rich and well suited for exploration using cold atom experiments.


\begin{acknowledgments}
We thank Debanjan Chowdhury, Joseph Thywissen, and Brian Demarco for helpful conversations.  This material is based upon work supported by the National Science Foundation under Grant No. PHY-1806357 and Grant No. PHY-2110250.
\end{acknowledgments}

\appendix

\section{\label{sec:eom}Equation of Motion for Entropy}

For a given distribution function, $f_k$, the Von Neumann entropy of the ensemble of fermions is
\begin{equation}\label{vn}
    S=-\int\frac{d^2k}{(2\pi)^2}\big(f_k\ln{f_k}+(1-f_k)\ln(1-f_k)\big).
\end{equation}
Near equilibrium, the distribution function has the form
\begin{equation}\label{linf}
f_k=f_k^0-\Phi_k \frac{\partial f_k^0}{\partial \epsilon_k},
\end{equation}
where $\Phi_k$ is  small.
We take the time derivative of Eq.~(\ref{vn}) and expand to leading order in $\Phi_k$:
\begin{equation}
\begin{split}
    \dot{S}&=-\int \frac{d^2k}{(2\pi)^2}~\ln\bigg(\frac{f_k}{1-f_k}\bigg)\dot{f}_k\\&\approx -\int \frac{d^2k}{(2\pi)^2}\big(-\beta(\epsilon_k-\mu)+\beta\Phi_k\big)\dot{f}_k.
    \label{eq:eEOM}
\end{split}
\end{equation}
We recognize Eq.~(\ref{eq:eEOM}) as an equation of motion for the total entropy, $\dot{S}=\beta\dot{E}-\beta\mu\dot{N}+\dot{S}_{neq}$, and conclude
\begin{equation}
\begin{split}
    \dot{S}_{neq}&=-\int \frac{d^2k}{(2\pi)^2}\beta\Phi_k \dot{f}_k.
    \label{eq:apEOM}
\end{split}
\end{equation}
Inserting $\dot{f}_k$ from
the linearized Boltzmann, Eq.~(\ref{eq:boltzeq}), into
Eq.~(\ref{eq:apEOM}) leads to the conclusion that
Eq.~(\ref{eq:entEOM}),
\begin{equation}
    \sum_i\xi_i\bigg(\frac{j^{(i)}_\alpha E_\alpha}{T}+u^{(i)}_\alpha\nabla_\alpha\bigg(\frac{1}{T}\bigg)\bigg)=\frac{1}{T}\sum_{ij}\xi_i\xi_jP_{ij}
    \nonumber,
\end{equation}
is equivalent to  $\dot{S}_{neq}=0$. 
The right hand side is the rate of entropy production from scattering processes.  This must equal the left hand side, the rate at which this heat is carried away.
In Appendix~\ref{sec:var}, we show that the optimal distribution function $f_k$ is obtained by maximizing the rate of entropy production.

\section{\label{sec:var}Variational Principle}
Following Ziman \cite{Ziman,Ziman2}, here we derive
a variational principle for transport coefficients.
We begin by introducing compact notation, defining
\begin{equation}
X_k=-\nabla_r
f^0_k\cdot v_k + eE\cdot\nabla_kf^0_k
\end{equation}
as the left hand side of the steady state Boltzmann equation, expanded to linear order in the electric field and thermal gradients.  We think of $X_k$ as components of a vector  and write $X$ as the abstract vector.  Similarly, $\Phi$ is the abstract vector with components $\Phi_k$ (see Eq.~(\ref{linf})).
We define the positive definite linear operator $P$ as
\begin{equation}
(P\Phi)_k = -I_k[\Phi],
\end{equation}
where $I_k[\Phi]$ is the linearized collision integral defined in Eq.~(\ref{eq:colInt}).  The linearized Boltzmann equation then reads
\begin{equation}\label{lbe}
X=P\Phi.
\end{equation}
We introduce an inner product,
\begin{equation}
    \langle \Phi,\Psi\rangle=\int\frac{d^2k}{(2\pi)^2}\Phi(k)\Psi(k).
\end{equation}
Taking the inner product of Eq.~(\ref{lbe}) with $\Phi$ yields
\begin{equation}\label{mom}
\langle \Phi,X\rangle=\langle\Phi,P\Phi\rangle,
\end{equation}
which can be recognized as the equation for entropy balance,
Eq.~(\ref{eq:entEOM}).

Let $\Phi$ be the exact solution to Eq.~(\ref{lbe}) and let $\Psi$ be a variational ansatz which obeys Eq.~(\ref{mom}), i.e. 
\begin{equation}\label{mompsi}
\langle \Psi,X\rangle=\langle\Psi,P\Psi\rangle.
\end{equation}
We will show that 
\begin{equation}\label{result}
\langle\Phi,P\Phi\rangle\geq
\langle\Psi,P\Psi\rangle,
\end{equation}
and hence the best variational solution is the one that maximizes $\langle\Psi,P\Psi\rangle$.  As argued in Appendix~\ref{sec:eom}, this corresponds to maximizing the entropy produced in collisions.

The proof is straightforward.  Since $P$ is positive definite, $\langle (\Psi-\Phi),P(\Psi-\Phi)\rangle\geq 0$.  Expanding this out yields
\begin{equation}\label{ineq1}
\langle \Phi,P \Phi\rangle\geq
-\langle \Psi,P \Psi\rangle +\langle \Psi,P \Phi\rangle +\langle \Phi,P \Psi\rangle.
\end{equation}
Explicitly writing out the integral reveals
$\langle\Phi,P\Psi\rangle=\langle \Psi,P\Phi\rangle$.  
We then use Eq.~(\ref{lbe}) and (\ref{mompsi}) to find
 $\langle \Psi,P\Phi\rangle = \langle\Psi,X\rangle=\langle\Psi,P\Psi\rangle$.  Substituting this into Eq.~(\ref{ineq1}) yields the desired result, Eq.~(\ref{result}).
 

\begin{widetext}
\section{\label{sec:lowTform}Collision Integral at Low Temperature}
Here we discuss the Sommerfeld expansion of the collision integral at low temperatures.
We will limit ourselves to $|\mu_F|\neq 0$, leaving the discussion of the half-filled case for Appendix~\ref{sec:tLinArg}. 
We will use a one-component ansatz, with $\phi_k=(\nabla_k \epsilon)_x=2t\sin k_x$.

Our starting point is Eq.~(\ref{eq:pij}).
We rewrite the energy and momentum delta functions as
\begin{equation}
    \delta(\epsilon_k+\epsilon_{k'}-\epsilon_{k''}-\epsilon_{k'''})=\int dE~\delta(\epsilon_k+\epsilon_{k'}-E)~\delta(\epsilon_{k''}+\epsilon_{k'''}-E)
\end{equation}
\begin{equation}
    \sum_Q\delta^{(2)}(k+k'-k''-k'''-Q)=\sum_Q\int d^2K~\delta^{(2)}(k+k'-K)~\delta^{(2)}(k''+k'''-(K-Q)).
\end{equation}
We now take the low-temperature limit of the product of Fermi functions, noting that both $f(\epsilon) f(E-\epsilon)$ and $(1-f(\epsilon))(1-f(E-\epsilon))=e^{\beta (E-\mu_F)} f(\epsilon) f(E-\epsilon)$ are sharply peaked about $\epsilon=E/2$, and that
\begin{eqnarray}
\int d\epsilon\, \frac{1}{e^{\beta (\epsilon-\mu_F)}+1}\frac{1}{e^{\beta (E-\epsilon+\mu_F)}+1}
&=&
\frac{E-2\mu_F}{1-e^{\beta (E-\mu_F)}}, 
\end{eqnarray}
which leads to the approximation
\begin{equation}
    f^0_kf^0_{k'}(1-f^0_{k''})(1-f^0_{k'''})\approx\frac{(E/2-\mu_F)^2}{\sinh^2(\beta(E/2-\mu_F))}~\delta(\epsilon_k-E/2)~\delta(\epsilon_{k''}-E/2).
    \label{eq:fermiDecomp}
\end{equation}
We substitute this leading behavior into Eq.~(\ref{eq:pij}), yielding a resistivity,
$\rho=P/j^2$, of the form
\begin{eqnarray}
\label{rho1}
\rho&=& \frac{\beta}{j^2} \int_{-8t}^{8t} \frac{(E/2-\mu_F)^2}{\sinh^2 \beta(E/2-\mu_F)} f(E) dE.
\end{eqnarray}
where $f(E)$  is an integral over the center of mass momenta of the colliding pairs that will be discussed below.  The current at zero temperature is simply $j(\mu_F)=2(e/\hbar)~\zeta(\mu_F/4t)$ where
\begin{equation}
\zeta(y)=\frac{4 |y|}{\pi^2}\bigg(E\big(1-y^{-2}\big)-\Pi\big(1+y^{-1},1-y^{-2}\big)-\Pi\big(1-y^{-1},1-y^{-2}\big)\bigg)
\end{equation}
and $E(k)$ and $\Pi(n,k)$ are complete elliptic integrals of the second and third kind, respectively. If $f$ is well behaved in Eq.~(\ref{rho1}), one can replace
\begin{equation}
\frac{(E/2-\mu_F)^2}{\sinh^2 \beta(E/2-\mu_F)}\to \frac{\pi^2}{3}T^3~\delta(E/2-\mu)
\label{eq:sinhExpand}
\end{equation}
which yields 
\begin{equation}\label{tsqrho}
\rho = \frac{\pi^2}{3}T^2 f(2\mu_F)/\big(j(\mu_F)\big)^2.
\end{equation}

The function $f(E)$ in Eq.~(\ref{rho1}) 
involves an integral over the incoming momenta $k,k^\prime$ and the outgoing momenta $k^{\prime\prime}$ and $k^{\prime\prime\prime}$. Due to momentum conservation, we can write $f(E)=\int d^2K~g(E,K)$, where $K$ is the center of mass momentum, and $g$ is an integral over the relative momenta. The only term in Eq.~(\ref{eq:pij}) coupling the incoming and outgoing integrals is the factor $(\phi_k+\phi_{k^\prime}-\phi_{k^{\prime\prime}}-\phi_{k^{\prime\prime\prime}})^2$.  Expanding this quadradic allows us to express $g$ as a sum of four terms, each of which are a product of incoming and outgoing terms,
\begin{equation}
    g(E,K)
    =\sum_Q\frac{4}{(2\pi)^5}\bigg( F^{(2)}(E,K)F^{(0)}(E,K-Q)+F^{(2)}(E,K-Q)F^{(0)}(E,K)-2F^{(1)}(E,K)F^{(1)}(E,K-Q) \bigg)
    \label{eq:Gk}
\end{equation}
where
\begin{equation}
    F^{(m)}(E,K)=2\sin \frac{K_x}{2}\int_{-\pi}^\pi d^2q~\cos^m (q_x)~\delta(\epsilon_{q+K/2}+\epsilon_{q-K/2}-E)~\delta(\epsilon_{q+K/2}-\epsilon_{q-K/2})
    \label{eq:Fm}
\end{equation}
and $K/2\pm q$ are the momenta of the two scattering particles.
Changing coordinates to $u=\cos(K_x/2)$ and $v=\cos(K_y/2)$, this can be rearranged to find
\begin{equation}
    f(E)=\frac{1}{4\pi^5}\int_{|E|/4}^1\frac{dv}{\sqrt{1-v^2}}\int_0^{v-|E|/4}\frac{du}{\sqrt{1-u^2}}\frac{v^2+u^2-2u^2v^2}{\sqrt{(u^2-v^2)^2(E^2/16-(u+v)^2)(E^2/16-(u-v)^2)}}.
    \label{eq:fullIntegral}
\end{equation}
We find empirically that this integral evaluates to
\begin{equation}
    f(E)=\frac{1}{16\pi^4}\sqrt{(4t/E)^2-1}.
    \label{eq:fullF}
\end{equation}
The physical consequences are discussed in Sec.~\ref{subsubsec:LowTRes}.

\section{\label{sec:tLinArg}Phase Space Integrals at Half Filling}
As presented, the integral in Eq.~(\ref{rho1}) is divergent due to the fact that $f(E)\propto E^{-1}$ for small $E$.  
This divergence is an artifact  of the approximation in 
Eq.~(\ref{eq:fermiDecomp})
where 
the product of Fermi functions is replaced with infinitely sharp delta-functions.  Here we show that at finite $T$ the divergence is cut-off, and as $T\to0$,  $f(E=0)\propto\beta$.

Including the finite widths of the Fermi function steps, Eq.~(\ref{eq:fermiDecomp}) takes on the form
\begin{equation}
    f^0_kf^0_{k'}(1-f^0_{k''})(1-f^0_{k'''})\approx\frac{(E/2-\mu)^2}{\sinh^2(\beta(E/2-\mu))}~\delta_\beta(\epsilon_k-E/2)~\delta_\beta(\epsilon_{k''}-E/2),
\end{equation}
where $\delta_\beta(x)$ has area $1$ and a width that scales as $1/\beta$.  The exact form is not important.
Setting $E=0$, the phase space integrals that appear in Eq.~(\ref{eq:Gk}) become
\begin{equation}
    F^{(m)}=2\sin\bigg(\frac{1}{2}K_x\bigg)\int_{-\pi}^\pi d^2k~\cos^m(k_x)~\delta(\epsilon_{k+K/2}+\epsilon_{k-K/2})~\delta_\beta(\epsilon_{k+K/2}-\epsilon_{k-K/2}).
    \label{eq:FmBeta}
\end{equation}
Note, the energy conserving delta-function is not broadened. For $E=0$, the function $F^{(1)}$ vanishes due to symmetry.

Along the diagonals ($K_x=\pm K_y$) 
the integrand is poorly behaved, and as $\beta\to\infty$ the integral is dominated by those regions.  To calculate the contribution from one diagonal, we shift the center-of-mass variables, $K_x=P+q$ and $K_y=P-q$, and consider the region where $|q|\ll P$.  The contribution from the other diagonals is identical.  


The functions $F^{(0)}$ and $F^{(2)}$ have the same scaling with $\beta$, so we only give the arguments for $F^{(0)}$. 
We use the energy conservation delta-function to perform the $k_y$ integral, treating $q$ as small,
\begin{equation}
    F^{(0)}(K;\beta)\sim\int dk_x~\frac{1}{|\sin(k_x)|}~\delta_\beta\bigg(4q\cos(P/2)\sin(|k_x|)-4q\frac{\sin^2(P/2)\cos^2(k_x)}{\cos(P/2)\sin(k_x)}\bigg).
    \label{eq:f0}
\end{equation}
There are now two small parameters in this problem ($1/\beta$ and $q$), so we must consider the asymptotic behavior of the integral for $\beta q\gg1$ and $\beta q\ll 1$ independently. In the former case, the broadened delta function is only nonzero when $k_x$ is within $\sim 1/|\beta q|$ of the points $P/2$ and $\pi-P/2$. The factor of $1/|\sin(k_x)|$ is well behaved in these regions, and we can replace it with $ 1/|\sin(P/2)|$.  
Treating $\delta_\beta$ as a box function, we see that the 
$\beta q\gg1$  contribution to the
integral 
scales
as $F^>\sim \beta(1/\beta q)^2$. The contribution to $f(E=0)$ from this region is then
\begin{equation}
f^> \sim \beta^2 \int_{1/\beta} \frac{dq}{q^4}\sim\beta.
\end{equation}
The $\beta q\ll 1$ contribution to the integral comes from the region where $k_x$ is not within 
$|\beta q|$ of the points $0$ and $\pm\pi$. The integrand diverges as $1/|k_x|$ near these points, so we need only consider the behavior in their vicinity:
\begin{equation}
F^<\sim
    \beta\int_{\beta q}\frac{dk_x}{|k_x|}\sim -\beta\ln(\beta q).
\end{equation}
The contribution to $f$ is has the same scaling
\begin{equation}
f^< \sim \int_0^{1/\beta} dq~\beta^2(\ln(\beta q))^2\sim \beta. 
\end{equation}
Thus we have established that the divergence is cut off, as described by Eq.~(\ref{eq:fT}).


\end{widetext}

\section{\label{sec:freeFermiSeebeck}Seebeck Coefficient of Non-Interacting Gas}
While some transport coefficients, such as the electrical and thermal resistivities, are undefined in the absence of scattering, the free Fermi gas has a well-defined Seebeck coefficient. The steady-state, collisionless Boltzmann equation  describes the behavior of the distribution function in response to electric fields and inhomogeneities:  Eq.~(\ref{eq:boltzeq}) with $\partial_tf_k=0$ and $I_k[f]=0$. As in Eq.~(\ref{eq:fkr}), we take $T(r)$ and $\mu(r)$ to be slowly varying, writing $f_k= f^0_k(r)$.  We absorb spatial derivatives of $\mu(r)$ into the  definition of the field, and hence the relevant spatial derivatives of 
 $f_k$ are proportional to $\nabla_rT$. We then take the moment of the Boltzmann equation with respect to $\nabla_k\epsilon_k$ to arrive at a steady-state condition for the particle number current:
\begin{equation}
    \nabla_rT\int\frac{d^2k}{(2\pi)^2}(\nabla_k\epsilon_k)^2\frac{\partial f_k}{\partial T}+ E\int \frac{d^2k}{(2\pi)^2}(\nabla_k\epsilon_k)^2\frac{\partial f_k}{\partial \epsilon_k}=0,
    \label{eq:freeSSBE}
\end{equation}
The Seebeck coefficient relates the electric field and thermal gradient, $E=\alpha \nabla_r T$, under the condition of a vanishing number current. We therefore rearrange Eq.~(\ref{eq:freeSSBE}) to find
\begin{equation}
    \alpha=\beta~\frac{\int\frac{d^2k}{(2\pi)^2}(\epsilon_k-\mu)(\nabla_k\epsilon_k)^2f^0_k(1-f^0_k)}{\int\frac{d^2k}{(2\pi)^2}(\nabla_k\epsilon_k)^2f^0_k(1-f^0_k)}.
\end{equation}
At high temperature, $T\rightarrow\infty$, the Seebeck coefficient approaches $\alpha\to-\beta\mu=\log(n_\sigma/(1-n_\sigma))$. At low temperature, $T\rightarrow 0$, the Seebeck coefficient vanishes linearly with temperature. 

\section{\label{sec:valid}Accuracy of Trial Functions}
Here we evaluate the accuracy of our variation trial wavefunction by systematically including higher moments.  We consider the low temperature limit, calculating $\rho$ via Eq.~(\ref{eq:thermDef}), including $N$ trial functions of the form $\phi^{(i)}=(\epsilon_k-\mu)^{i-1}(\nabla_k\epsilon_k)_x$.
At low temperatures, the scattering integrals $P_{ij}$ defined in Eq.~(\ref{eq:pij}) can be expanded as shown in Appendix~\ref{sec:lowTform}. In particular, using the approximation in Eq.~(\ref{eq:fermiDecomp}) the low-temperature expression for $P_{ij}$ is
\begin{equation}
    P_{ij}=\beta\int_{-8t}^{8t}\frac{\big(E/2-\mu_F\big)^{i+j}}{\sinh^2{\beta(E/2-\mu_F)}}f_T(E)dE,
\end{equation}
where the function $f_T(E)$ is defined in Eq.~(\ref{eq:fT}). We then expand the integrand using Eq.~(\ref{eq:sinhExpand}) to determine the leading-order low-temperature behavior of $P_{ij}$. The currents $j^{(i)}$ and $u^{(i)}$ (see Eq.~(\ref{eq:currents})) are expanded in an analogous manner, and we determine the thermoelectric matrix using Eq.~(\ref{eq:thermoelectricMat}).


We define $\rho_N$ as the resistivity calculated using all trial function $\phi^{(i)}$ with $i\leq N$. We find that including terms beyond $n=1$ simply rescales the thermoelectric response functions: at low temperatures, the ratios between different approximants, $\rho_N/\rho_1$, are temperature and density independent.

\begin{figure}[tbph]
    \centering
    \includegraphics[width=3.375in]{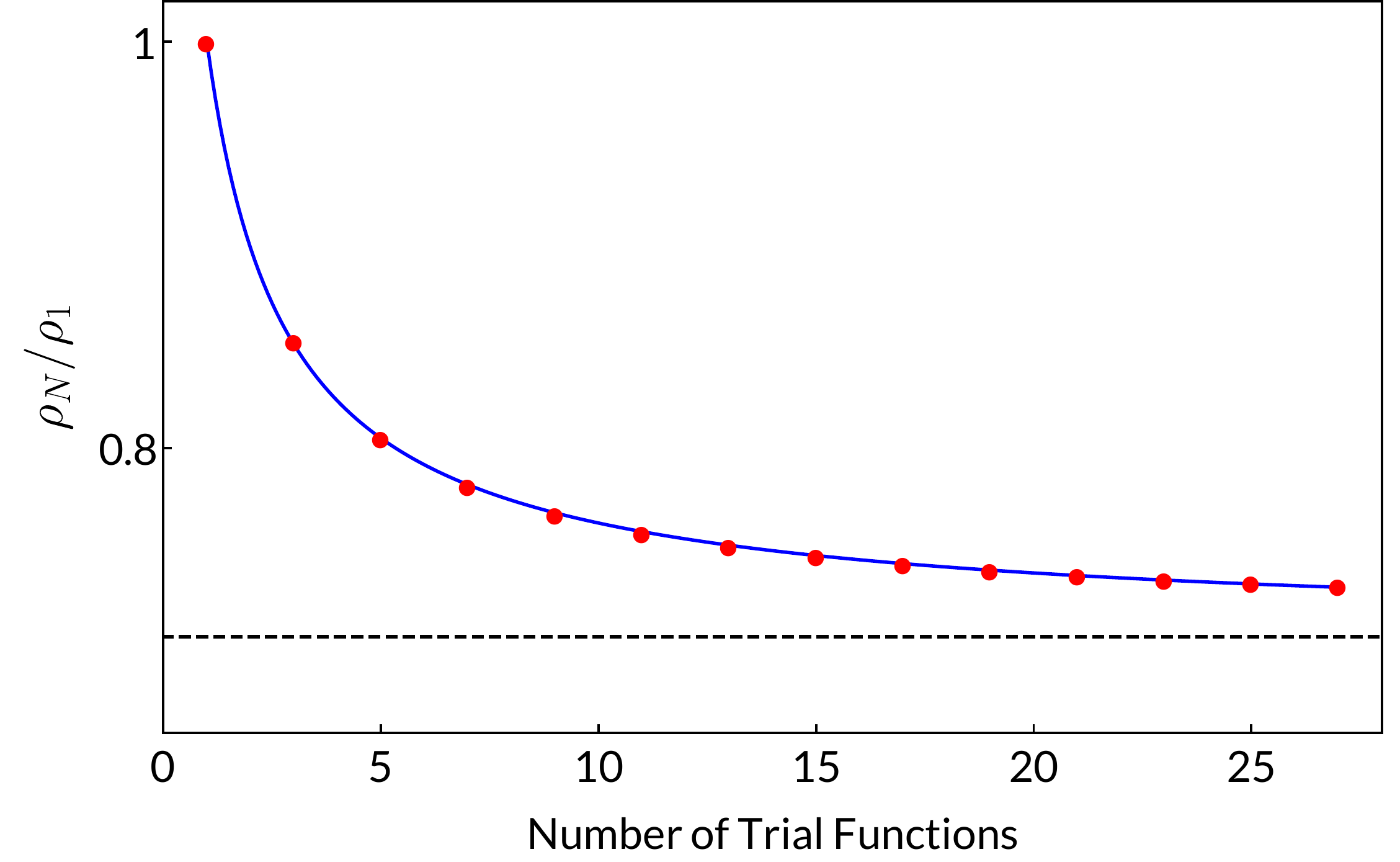}
    \caption{Low-temperature resistivity  $\rho_N$ calculated using a variational ansatz using $N$ trial functions of the form
    $\phi^{(i)}=(\epsilon_k-\mu)^{i-1}(\nabla_k\epsilon_k)_x$.  At low temperature including more terms simply rescales the resistivity, and  the ratio $\rho_N/\rho_1$ is independent of all microscopic parameters. 
    Blue line:  Best fit of the form $\rho_N/\rho_1=1-A\sum_{i=1}^N i^{-\alpha}$,
    with $\alpha=1.86$ and $A=0.368$. Black dashed line: Asymptote of the fitting curve at $\rho_\infty/\rho_1=0.707$.}
    \label{fig:moreTerms}
\end{figure}

Figure~\ref{fig:moreTerms} shows how the resistivity changes as we add more terms to our ansatz.
The calculation is variational, so the resistivity monotonically decreases as more terms are added.  Extrapolating $N\to\infty$
gives a $30\%$ reduction from the $N=2$ result discussed in the main paper.  More general ansatze are unlikely to significantly change this result.
Similarly,
it is reasonable to 
assume that this estimate of the error applies at all temperatures.




\providecommand{\noopsort}[1]{}\providecommand{\singleletter}[1]{#1}%

\end{document}